# Precise determination of reliable work function in according to the definition from photoelectric effect

Liu Changshi

Nan Hu College, Jiaxing University, Zhejiang, 314001, P. R. China，Tel.:086-15325357233, Fax. : 086-0573-83643085, E-mail: lcswlx@163.com

Abstract

The strategy of using the definition of work function from photoelectric effect to determine precise and reliable work function by the contribution of frequency to photoelectric yield is investigated theoretically in this paper. Based on the Fermi-Dirac distribution and the definition of work function proposed by Einstein in photoelectric effect, a typical numerical method is pursed step by step to quantitatively analyze the frequency-dependent photoelectric yield. Supplementing applications to $In_{14}$ cluster and three kinds of metals, the simulations agree well with the observed spectra (photoelectric yield-frequency). At the same time, the threshold frequency of light-dependent the work function is theoretically explained successfully via one equation so that the work function can be predicted precisely and reliably. These results suggest that the formalism pursed in this paper, which is straight forward and physical, may be of significant utility in metal cluster spectroscopy. It is hoped also that the results will encourage a comprehensive theoretical analysis of the applicability of bulk-derived models to cluster photo-ionization behavior, and of the transition from atomic and molecular-type to surface-type photoemission.

1. Introduction

Light not only can produced Photoelectric Effect on bulk metal surfaces [1-5], but also can created Photoelectric Effect in metal cluster [6]. Based on the explanation on photoelectric effect of Einstein [7], the work function can be detected after the curve of frequency-depend stopping potential is fitted. Work function is one key to metal cluster ionization potentials as an important characteristics of these "artificial atoms." However, a difficulty in this method is that the equivalent of the stopping potential is not sharp: As the retarding potential approaches stopping potential, the anode current goes asymptotically to a constant value. This is partly because the electrons in the photocathode have a thermal energy distribution which broadens the work function. Hence, the stopping potential is detected by extrapolating with geometry feature so that the reliability of the stopping potential is compromised. Another of the most widely applied method of determining work function was established which is the thermion electron emission from the cathode surface is to day known as the Richardson-Dushman equation [8]. But it should be noted that the Richardson-Dushman equation was developed via classical statistics named Maxwell-Boltzmann instead of quantum statistics, so the reliability of the Richardson-Dushman equation is not the highest. Although there has been Fowler`s theory of photoelectric emission or Fowler formula [9], Fowler formula was concerned with values of hv near the work function and very low temperature. [9]

The interest in work function is growing in the context of metal clusters [6], new metal, new electrode and new alloy [10-19]. Studying functional metal requires the most accuracy knowledge of the work function of the metal. In view of this requirement, an attempt is made to predict the work function of the metal more precise and more reliable via natural path in this paper. Physical experience leads people to believe that specification function on experimental results defines a unique physical problem，even if the specification function is empirical functional forms. In order to obtain the work function more precisely and reliably, based on the Fermi-Dirac distribution and photoelectric equation of Einstein, a natural path and formulation is suggested in this paper. This method may be the most precise determination of work function.

2 Methodologies

Because electron is fermions, the Fermi-Dirac distribution should apply to electron. Hence, Fermi-Dirac distribution can describe the photoelectrons in an external electronic field [8, 9], The probability of an electron state with total energy E being occupied is given by the Fermi-Dirac (F-D) distribution function

$$f(E) = \frac{1}{1 + \exp((E - E_f)/kT)} \qquad (1)$$

where f(E) is the probability that a particle will have energy E, $E_f$ is Fermi energy, T is absolute temperature, k is Boltzmann constant and its value is $k = 1.38 \times 10^{-23}$ in SI unit.

The maximum kinetic energy of photoelectron is $eV + h\nu - h\nu_0 - e\delta$ after it has been liberated from the metal (external photoelectric effect) and enter into the external electronic field with V. Where e=electron charge=$1.602 \times 10^{-19}$, $h\nu_0$ is the work function of the metal. $h\nu_0$ is the energy you must give to an electron at the Fermi level to kick it out of the metal and turn it into a free electron. An electron below the Fermi level needs more than $h\nu_0$ to escape, δ is an additional contact potential because the surfaces of the anode and cathode are different. An electron absorbs a photon and now has the maximum photoelectron kinetic energy $eV + h\nu - h\nu_0 - e\delta$, thus the probability that an electron will have the maximum photoelectron kinetic energy $eV + h\nu - h\nu_0 - e\delta$ is

$$f(E) = \frac{1}{1 + \exp((eV + h\nu - h\nu_0 - e\delta)\frac{1}{kT})} \quad (2)$$

A change of equation (2) variable to $\nu$ yields

$$f(E) = \frac{1}{1 + \exp(\frac{\nu - \nu_1}{\nu_2})} \quad (3)$$

where $\nu_1$ is $\frac{1}{h}(h\nu_0 + e\delta - eV)$ and $\nu_2$ is $\frac{kT}{h}$. One natural conclusion is that photoelectric yield, Y(ν), defined as photo-current by per unit light intensity, is proportional to the probability, f(E), that a electron will have energy $eV + h\nu - h\nu_0 - e\delta$, in mathematical language:

$$Y(\nu) = \frac{Y_2}{1 + \exp(\frac{\nu - \nu_1}{\nu_2})} \quad (4)$$

But mathematical result shown that:

$$\frac{1}{1 + \exp(\frac{\nu - \nu_1}{\nu_2})}\bigg|_{\nu \to \infty} = 0 \quad (5)$$

Experimental results of photoelectric effect tell that as the frequency of incident light approaches high enough, the anode yield, Y(ν), goes asymptotically to a constant value know as the maximum yield, $Y_{max}$:

$$Y(v)|_{v \to \infty} = Y_{max} \qquad (6)$$

Boundary condition (6) predicts that the photoelectric yield and the frequency of incident light should be related as:

$$Y(v) = Y_1 + \frac{Y_2}{1 + \exp(\frac{v - v_1}{v_2})} \qquad (7)$$

Combining equations (5), (6) and (7) yields

$$Y_1 = Y_{max} \qquad (8)$$

Substitution of equations (8) into (7) leads to following expressions:

$$Y(v) = Y_{max} + \frac{Y_2}{1 + \exp(\frac{v - v_1}{v_2})} \qquad (9)$$

Where $Y_2$ is the yield constant expected by Equation (9), $Y_{max}$ is the maximum yield predicted via Equation (9), $v_1$ is the inflection point frequency, when the frequency is lower than $v_1$, the Y–v curve is concave, while, as long as the frequency is higher than $v_1$, the Y–v curve is large convex, meanwhile, the average value of the photoelectric yield is fixed at $v_1$, the $v_2$ is constant of frequency. $Y_2$, $Y_{max}$, $v_1$ and $v_2$ will be optimized. Hence, the equation (9) is the quantitative relationship between photoelectric yield and frequency.

To find the threshold frequency $v_0$, the function (9) should be equated to zero because photoelectrons can not be observed only if $v \leq v_0$ in according with the photoelectric theory of Einstein.

$$Y(v_0) = 0 = Y_{max} + \frac{Y_2}{1+\exp(\frac{v_0-v_1}{v_2})} \tag{10}$$

Solution equation gives the value of the threshold frequency:

$$v_0 = v_1 + v_2 Ln[-(1+\frac{Y_2}{Y_{max}})] \tag{11}$$

It can be stated that a method based on simulation of the data of frequency-dependent photoelectric yield, rather than by extrapolating method is developed this manuscript.

## 3. Application

The measurement of the light energy-yield of $In_{14}$ cluster is used to justify the accurate application of equation (11). FIG. 1 shows the measured normalization photoelectric yield of $In_{14}$ cluster versus photon energy [6]. This paper determines the photoelectric yield according to the above approach (9). Then through experimental raw data employed in the method of the regression analysis, the experimental curves for normalization photoelectric yield along energy have been simulated using the component spectrum shown in Fig. 1. Optimized parameters employed to simulate the component spectrum are also listed in Table 1. The modeled verification is also carried out by a comparison of modeled photoelectric yield obtained by numerical integration of photoelectric yield to measured photoelectric yield in Fig. 1.

Although the method suggested in this paper has been found to apply successfully to In$_{14}$ cluster, this does not give enough idea about natural method and more accuracy of the methodology suggested in this paper. To test the here in the determination of the work function of any kind of metals using experimental data of the curve constructed by frequency-photoelectric yield, the experimental photoelectric yields of Molybdenum [20], Palladium [21] and potassium on a thick layer of oxygen on tungsten in electrostatic fields of 0 Vm$^{-1}$ after being heated to 1873.15 K [22] are illustrated in Fig. 2, 3 and 4. Of course, the plots of photoelectric yield versus frequency have been fitted with equation (9) and shown as cross in Fig. 2, 3 and 4.

The solution of equation (11) reveal a work function of 4.30 eV for Molybdenum, the work function of Molybdenum reported via Fowler formula is 4.15 eV [20]. The best parameters to obtain the best results of calculation of Palladium are listed in Table 1. When the equation (11) is applied in Table 1, it is found that the work function of Palladium is to be 4.85±0.17 eV, the work function of Palladium was detected to be 4.99±0.04 eV by Fowler formula[21]. The work functions of potassium reported by both frequency-depend stopping potential [22] and Richardson-Dushman equation[22] is very close to the work function obtained in this paper, the form is 2.24 eV and the latter is 1.9 eV.

In order to verify the accuracy of the simulation, the correlation coefficient between the measured and the simulated data is given in Table 1; the minimum magnitude of the correlation coefficient is 0.989. The average relative error (ARE) ($\frac{1}{n}\sum_{i=1}^{i=n}\frac{|Y_{im} - Y_{is}|}{Y_{im}} \times 100\%$) to evaluate the simulation results is also shown in Table 1, where $Y_{im}$ is measured data, and $Y_{is}$ stands for the value obtained by simulation. The maximum value of ARE is 9.27%. As is shown in Table 1, FIG. 1, 2, 3 and 4, it can be stated that a satisfactory agreement between the measured and the modeled Y is achieved by the functions given in Table 1.

The solution of equation (11) reduces the work function by the methodology proposed in this paper is a numerical calculation path. It is evident from FIG. 1 to 4 and Table 1 that the contrary to the frequency-depend stopping potential with extrapolating, one can confirmed that the work function calculated by the natural method proposed in this paper is independent of the value of stopping potential, as the best benefit the work function is predicted with very high reliability. Because almost each pair of data constructed by frequency and photoelectric yield contribution to the prediction on the work function, the work function could thus be predicted with considerable accuracy by photoelectric emission.

4. Conclusions

In order to obtain a new method for predicting the work function through natural path and more precise, the photoelectric yield was calculated as a function of frequency of light, firstly. A typical mathematical function was suggested. The function can describe the change trend of the photoelectric yield with frequency across the measured range. The theoretical values simulated in this paper are consistent with well experimental data, the maximum data of mean relative errors is 9.27% and the minimum value of the correlation coefficient between the actual and the calculated data is 0.989. Based on the photoelectric equation of Einstein, a numerical equation has been developed to determine the work function more precisely and reliably because there is the function to link photoelectric yield and frequency of light. Application results state that curve consisted of photoelectric yield and frequency of light for a given material is just sufficient to calculate its work function, this method of determination of the work function is independent of an exact measurements of the frequency threshold, and depends only on the measurement of the photoelectric yield along the frequency of light. All parameters applied in this model are not only easy to seek, but also the scientific significance of them can be discovered during using the method. One can find that this method may be the most accurate, easiest and fastest for work function determination.

Work-Functions and Photo-Electric Thresholds of the Alkali Metals, Proc. Roy. Soc., A., 1925, 107(743) , 377-410,. ,

Table 1 Results and justification of simulation to obtain the work function and evaluations for the photoelectric yield of $In_{14}$ cluster and three kinds of bulk metal as a function of the frequency of the incident light.

| Materials | Function form | R | ARE (%) | $\nu_0$ ($\times 10^{15}$ Hz) | $h\nu_0$ (eV) |
|---|---|---|---|---|---|
| $In_{14}$ cluster | $Y(E) = 1.50 - \dfrac{1.57}{1 + \exp((E - 6.05)/0.22)}$ | 0.989 | 9.27 | | 5.39 |
| Molybdenum | $I(\nu) = 958.72 - \dfrac{6.95 + 958.72}{1 + \exp((\nu - 1.49)/0.09)}$ | 0.999 | 6.98 | 1.04 | 4.30 |
| Palladium, 1078 K | $Y(\nu) = 7.75 - \dfrac{7.67}{1 + \exp((\nu - 1.29)/0.027)}$ | 1.0 | 0.94 | 1.2 | 4.99 |
| Palladium, 730 K | $Y(\nu) = 9.31 - \dfrac{9.55}{1 + \exp((\nu - 1.30)/0.031)}$ | 0.999 | 0.56 | 1.12 | 4.66 |
| Palladium, 305 K | $Y(\nu) = 8.82 - \dfrac{9.08}{1 + \exp((\nu - 1.30)/0.027)}$ | 0.999 | 8.43 | 1.17 | 4.91 |
| Potassium | $I(\nu) = 150.57 - \dfrac{151.57}{1 + \exp((\nu - 6.03)/0.13)}$ | 0.999 | 0.12 | 5.38 | 2.22 |

R: Correlation Coefficient, ARE: Average Relative Error

Figure caption

Fig. 1 Numerical results of the photoelectric yield as a function of the energy of the incident light for $In_{14}$ cluster.

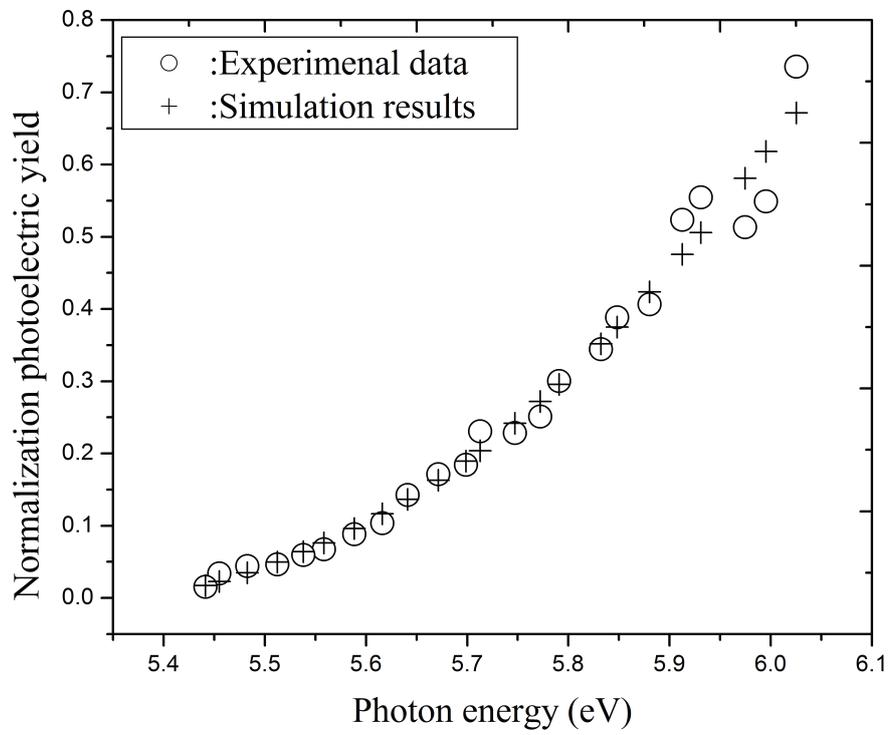

Fig. 2 Comparisons between experimental and calculated frequency dependence of the photoelectric yield of Molybdenum.

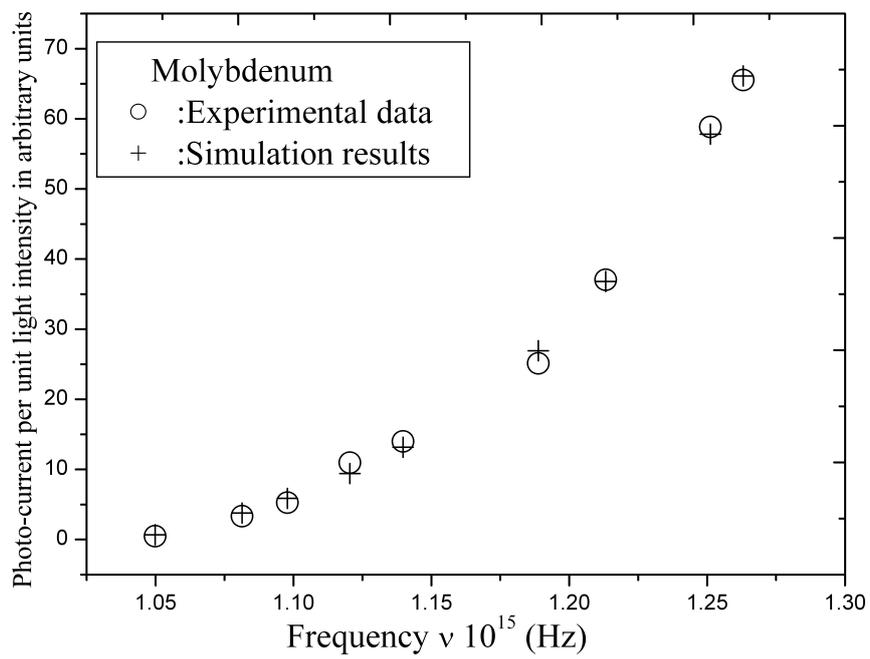

Fig. 3 Geometry used to determine the photoelectric yield of Palladium at various temperatures by the frequency of the light.

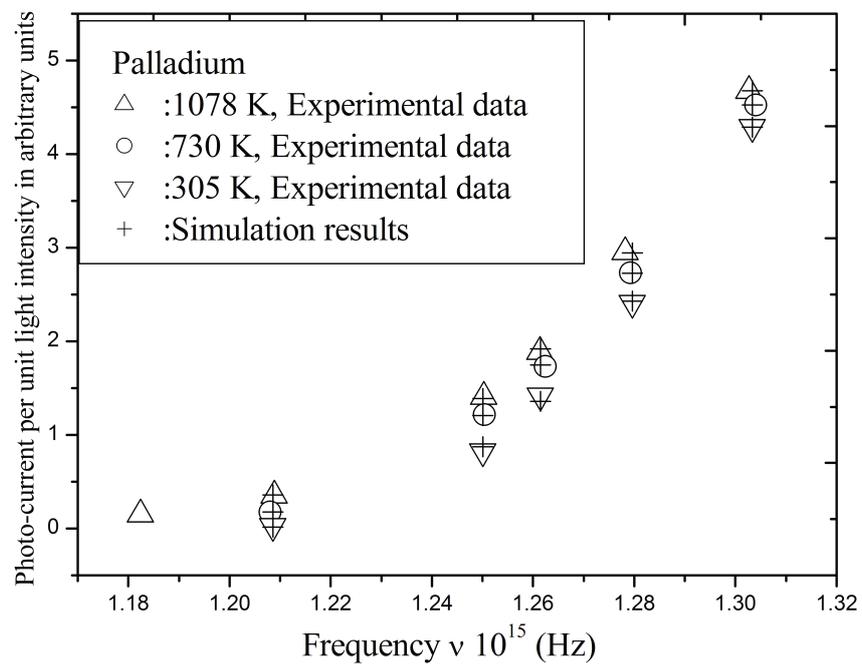

Fig. 4 Scatter plots of both the photoelectric yield expected by Equation (9) against the frequency of light and the experimental results of the potassium on a thick layer of oxygen on tungsten in electrostatic fields of 0 Vm$^{-1}$ after being heated to 1873.15 K.

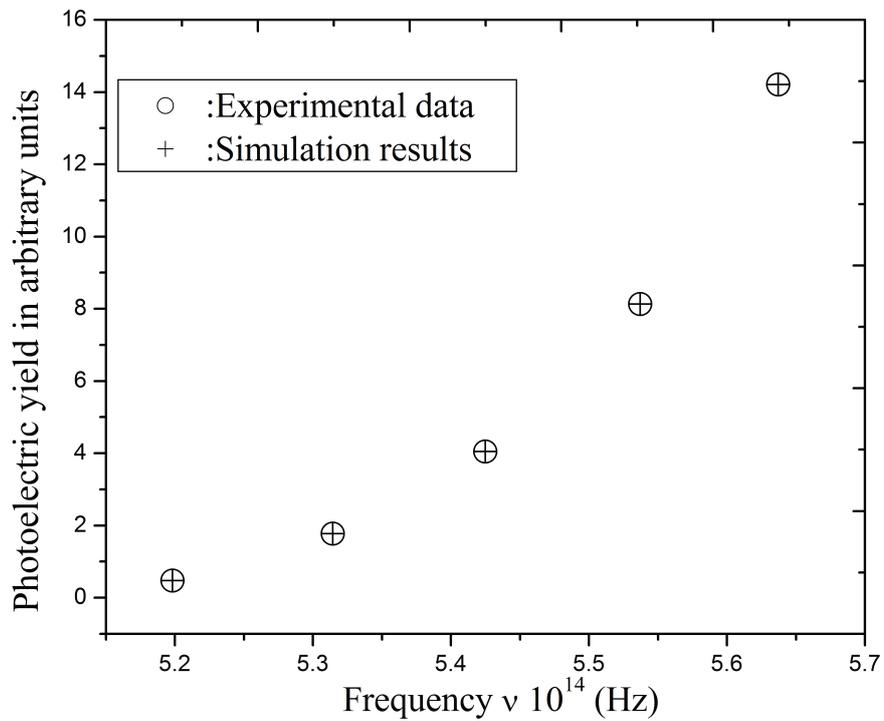